\documentclass[a4paper]{article}
\usepackage[utf8]{inputenc} 
\usepackage[affil -it]{authblk}
\usepackage{graphicx}
\usepackage{verbatim}
\usepackage{amsmath,amsfonts,amssymb,textcomp}
\usepackage{color}
\usepackage{caption}
\usepackage{subcaption}
\usepackage{epstopdf}
\usepackage{hyperref}

\begin{document}
\title{Electron scattering on hydrogen and deuterium molecules at 14-25 keV by the "Troitsk nu-mass" experiment}
\author[1]{D.\,N.\,Abdurashitov}
\author[1]{A.\,I.\,Belesev}
\author[1]{ V.\,G.~Chernov}
\author[1]{ E.\,V.\,Geraskin}
\author[1]{A.\,A.\,Golubev}
\author[1,2]{G.\,A.~Koroteev}
\author[1]{N.\,A.~Likhovid}
\author[1,2]{A.\,A.~Nozik}
\author[1]{V.\,S.\,Pantuev}
\author[1]{V\,.I.~Parfenov}
\author[1]{A.\,K.~Skasyrskaya}
\author[1]{S.\,V.~Zadorozhny}

\affil[1]{Institute for Nuclear Research of Russian Academy of Sciences, prospekt 60-letiya Oktyabrya 7a, Moscow 117312, Russian Federation}

\affil[2]{Moscow Institute of Physics and Technology, 9 Institutskiy per., Dolgoprudny, Moscow Region, 141700, Russian Federation}

\maketitle

\begin{abstract}

We've performed precise measurements  of electron scattering on molecular hydrogen and deuterium by using the "Troitsk nu-mass" setup. Electrons were generated by the electron gun with an energy line width better than 0.3 eV. The electron energies were 14, 17, 18.7, 19 and 25 keV. The windowless gaseous  tritium source (WGTS) was filled by hydrogen isotopes and served as a target. The total column  density was adjusted to form a length of 0.35--0.7 of the  electron mean free path. The integral spectrum of scattered electrons was measured by the electrostatic spectrometer with a magnetic adiabatic collimation and relative energy resolution 8.3$\cdot 10^{-5}$. As a result, the shapes of molecular excitation and ionization spectra were extracted for both isotopes. We did not find any difference between hydrogen and deuterium targets. The relative energy dependence was extracted too.

\end{abstract}

\section{Introduction}

Experimental measurements of electron cross-section at keV energy with atomic and molecular hydrogen are relevant for  astrophysics and physics of low-temperature plasma~\cite{drawin1976instantaneous}.  Intergalactic space electrons with low energy play an important role   in the investigation of interstellar clouds evolution~\cite{Padovani:2009bj}. 
Relation between molecular excitation and ionization cross-sections is also crucial in calculation of electron trapping in the system with magnetic mirrors. Such a trap is formed by magnetic field configuration in the  windowless gaseous  tritium source (WGTS) used by experiment "Troitsk nu-mass" in search for a sterile neutrino sign in the tritium $\beta$-decay spectrum~\cite{Abdurashitov:2015jha}. This experiment is aimed to measure  $\beta$-spectrum in a wide energy range. Thus, understanding of the trapping effect at different electron energies becomes critical too. The previous measurements of electron scattering on molecular tritium in the Troitsk experiment were carried out at a fixed electron energy of  18.6 keV~\cite{aseev2000energy}  (in that article one could also find the energy loss measurement in quench condensed $D_2$ films). The energy resolution in that measurement was about 3.5~eV. In the current setup we have a new spectrometer with an energy resolution of 1.5 eV for 19 keV electrons~\cite{Abdurashitov:2015jha}. This allows us to measure the molecular excitation peak with better precision. 

Accurate descriptions of inelastic cross-section for electron collisions with light molecules and atoms were published long  ago~\cite{lotz1967empirical,miller1957theory}. For energies of electrons much higher than the molecular ionization potential the Born approximation is applied to $H_2$ with a wave function with two interaction centers. This method allows one to evaluate cross-sections for all possible processes: generation of negative and positive molecular ions, excitation  of molecular states~\cite{bartschat1989excitation}, dissociation or their combination~\cite{liu1973total}. 

The most practical formula for hydrogen matching the experimental cross-section was  presented by Lotz~\cite{lotz1968electron}:  

\begin{equation}    \label{Lotz}
	\sigma_{ion} \sim \frac{\log\left(\frac{E}{I} \right)}{E \cdot I}\left[ 1- b\exp \left(-c\left( \frac{E}{I}-1 \right) \right) \right]
\end{equation}
where $\sigma_{ion}$ - ionization cross-section, $E$ -energy of electron, $I$ - ionization potential, $b, c $ - numerical constants. The expression in square brackets describes behavior at the electron energy of a few hundred electron-volts. At the energy much higher than the ionization potential $I$ formula~\ref{Lotz} simplifies to 

\begin{equation}    \label{eq:energy}
	\sigma_{ion} \sim \log \left(\frac{E}{I} \right)/(E\cdot I).
\end{equation}
In paper~\cite{schram1965ionization} the absolute ionization cross-sections for different gases  were measured at energies under 20~keV. Our interest is not only in ionization cross section and its energy dependence but in total inelastic cross-section including molecular excitation and relation between excitation and ionization probabilities. 

In this paper we describe measurements of electron scattering on gaseous hydrogen and deuterium using the electron gun and the new spectrometer. The shapes of excitation and ionization spectra were extracted for electron energy of 14, 17, 18.7, 19 and 25 keV. The paper is organized as a follows: first, the Troitsk experiment set up is briefly described, then follows the explanation of measurement procedure and data analysis. The last chapters presents our results and summary. 

\section{Experimental set-up and procedure}

The experimental set-up was very identical to that used in the previous measurement~\cite{aseev2000energy} but with a new spectrometer which has about a factor of two better energy resolution~\cite{Abdurashitov:2015jha}. In brief, a Windowless Gaseous Tritium Source (WGTS) filled with hydrogen isotopes served as a target. the WGTS with an axial magnetic field was filled with gas, circulating in a closed loop. The main part of the gas being exposed to the electron beam was contained in a 3 meter long pipe. The pipe was kept at a constant temperature of about 30~K which is above the boiling point for hydrogen and deuterium but lower than freezing temperature for nitrogen and oxygen (thus effectively removing these gases from interaction volume). The pipe was continuously cooled down by the gaseous helium flow returning from cryostat to the refrigeration system, and heated by the heaters wrapped around the pipe. The pipe temperature was controlled by a set of sensors alongside the interaction volume and stayed better than $\pm0.3$K during one data taking run. This defines the relative gas density stability at the level of $0.3/30=0.01$. The average column density of hydrogen isotope was chosen to be about  0.35--0.7 of the electron mean free path depending on electron energy and a particular fill which corresponds to about $5\cdot 10^{16}$ molecules per cm$^2$ . Electrons were generated by the electron gun located at the rear side of the WGTS. The gun has a half transparent gold-plated quartz cathode connected to the negative potential power supply and is illuminated by ultraviolet deuterium lamp. Details of the gun design can be found in~\cite{Abdurashitov:2015jha}. The electron energy spread is defined by stability of the high voltage power supply (HV), it was found to be less than $\pm$0.3~eV. Measurements were taken in two separate periods: at electron energy 14, 18.7 and 25 keV in the first run and at 14, 17, 19 keV in the next one. 

\begin{figure}[htb]
\center
	\includegraphics[width=1.\linewidth]{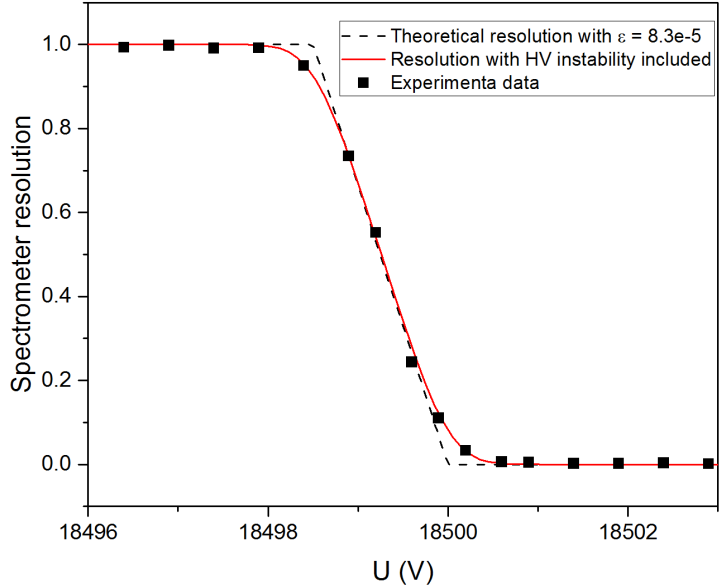} 
	\caption{Spectrometer resolution function measured for 18500 eV electrons versus potential on the electrostatic electrode.}
	\label{fig:resol}
\end{figure}

WGTS has a differentiating pumping system which prevents gas penetration into the spectrometer. Electrons from the gun pass WGTS, scatter there with some probability on target gas then enter the spectrometer with high vacuum and a strong magnetic field, $B_0=7.2$ Tesla, in the pinch magnet and then are adiabatically guided into a low field $B_m=6\cdot10^{-6}$ Tesla in the center of the spectrometer. The size of the spectrometer  is about 10 meter in length and 2.8 meter in diameter. At the same time electrons are de-accelerated by a strong electrostatic field formed by a single cylinder-shaped electrode sitting at the analyzing potential~\cite{Abdurashitov:2015jha}. The spectrometer serves as an electrostatic filter with a magnetic adiabatic collimation (MAC-E filter). By its principle this is an integral spectrometer. The filter transmission function has a step-like shape versus the applied potential and defines the slope versus the electron energy as $\Delta E = E \cdot B_m/B_0$, Fig.~\ref{fig:resol}. At the other side of the spectrometer there is another superconducting solenoid magnet with a field strength of 2.1 Tesla, which collects all electrons onto the detector at the ground potential. The electrons reaching the detector are accelerated to the original energy after passing the electrode. 

\begin{figure}[htb]
\center
 	\includegraphics[width=0.8\linewidth]{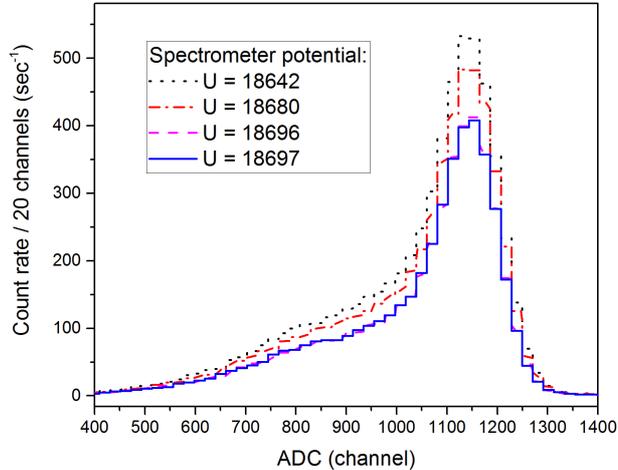}
    \caption{Amplitude spectrum from Si-detector during electron scattering measurements at 18700 eV in hydrogen at different  HV spectrometer potentials. Two lower histograms correspond to the events without scattering, see flat region in Fig.~\ref{fig:scat}, and the two others -- for points of integral spectrum of the scattered and non scattered electrons.}
	\label{fig:chanels}
\end{figure}

The Si (Li)-detector with a 10~nm gold window detects electrons with the energy above 2 keV. After selecting a proper window for detector signal amplitude, see Fig.~\ref{fig:chanels}, the relative counting rate at a particular spectrometer potential was the main source of information for the scattering events.  

The measuring procedure was the following: after getting all superconducting magnets at their nominal currents and confirming that the WGTS pipe temperature was stabilized, some portion of gas was injected into the closed gas loop. The relative amount of gas was controlled by a pressure sensor ($P_x$) located in the  warm (room temperature) part of the gas loop. $P_x$ is positioned between the pump, which collects gas from both sides of the pipe, and the long nozzle which cools gas to the pipe temperature and then injects gas in the middle of the pipe. A typical value of pressure measured by $P_x$ in this part of the gas flow  loop was around 2.8--3.2 mm of Hg. We have to emphasize that because of the nozzle at the entrance to the pipe, gas concentration inside the pipe was much smaller than the value of $P_x$. At the same time, it was found that gas concentration in the pipe and $P_x$ are related very linearly. Beside $P_x$ gas pressure was controlled by vacuum meters on both sides of the WGTS. At the rear side a mass analyzer was connected to the pipe, which controlled gas composition. 

The electron energy from the gun was set by a high voltage (HV) power supply  between -14~kV and -25~kV. The output from this HV supply was also connected to another $\Delta$HV module, whose output gives an additional $\Delta$HV potential. This $\Delta$HV can vary in the range -20~V to +900~V with a step of 0.1~V on top of the value of the main HVpower supply. The output of the $\Delta$HV module is connected to the spectrometer. Such a configuration prevents possible high voltage fluctuations in case of using two absolutely independent HV modules. The absolute values of both power supplies were controlled by two independent resistor dividers. The HV and dividers in sum gave stability better than 0.15V over 200 seconds periods.     

\begin{figure}[htb]
\center
	
	\includegraphics[width=0.8\linewidth]{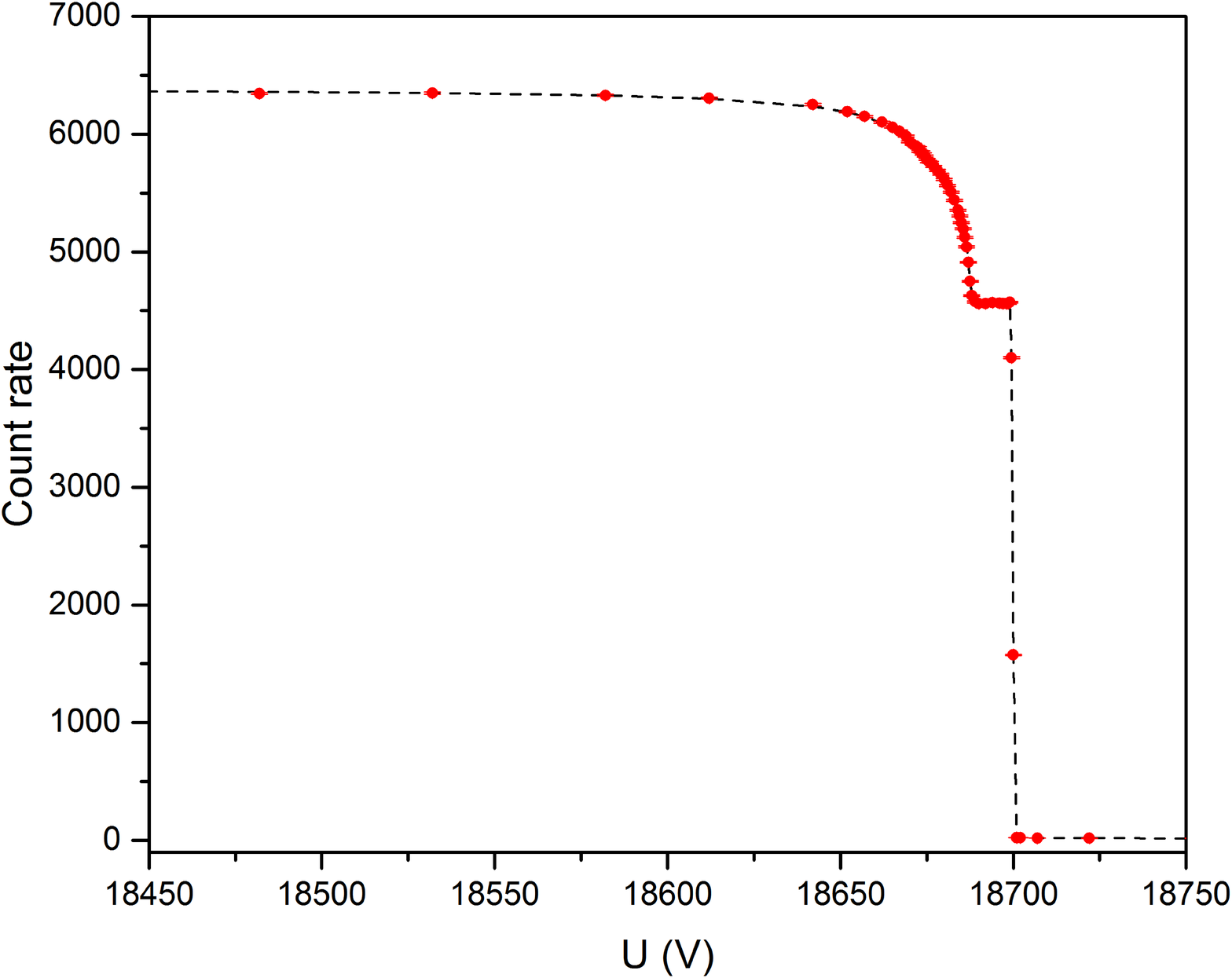} 
	\caption{ Typical integral spectrum measured for 18700 eV electron scattering in hydrogen.}
	\label{fig:scat}
\end{figure}

The goal of the measuring procedure is to scan electrostatic potential of the spectrometer in the range slightly above the electron energy from the gun to the value down to 300~V to catch the spectrum of the scattered electrons. A typical integral spectrum measured at 18700~eV is shown in Fig.~\ref{fig:scat}.
There is a sharp edge on the right similar to the step shown in Fig.~\ref{fig:resol}, the slope of which is determined by the energy resolution. Then, to the left from the step there is a flat area of about 12~V width which corresponds to the energy gap between the non-scattered electrons and the electrons which lost energy after molecular hydrogen excitation and ionization. The hight of this step relative to the rate at the spectrometer HV lower than the electron energy by 200--300~V (left points in Fig.\ref{fig:scat}) roughly defines the probability for electron to pass WGTS without scatterings. Then the integral spectrum goes up reflecting the number of electrons which lost some energy, see below. The electron intensity was chosen to lie between 3--6~kHz due to dead time of DAQ. 

The whole spectrum was measured over 65--74 set points. To avoid the influence of a possible intensity drift of the deuterium lamp (and consequently, the electron flux) control measurements were made at a monitor point of 300~V down from the gun potential. Monitor points were collected after every two set points. The counting rates between two monitor points were corrected for intensity drift if this monitor distortion exceeded two statistical errors. Typically, intensity was stable within 0.5\% between two monitor points. Overall, it takes about half an hour to write one data file. For one gas fill in WGTS, for each electron energy set, we try to keep even number of files by going in forward and reverse high voltage directions. In the second round of measurements when the electron energy was 14, 17 and 19 keV we also measured two energy points for each gas fill: 14~keV and 19~keV or 17~keV and 19~keV. This allowed us to minimize the influence of different column density from fill to fill and to get a more precise cross section energy dependence.

\section{Data processing and analysis} \label{section:analysis}

\subsection{Preliminary data processing}

The preliminary data analysis for this experiment was done in the following sequence:
		
\begin{enumerate}
	\item Initial readout of binary data files. The files contain a set of measurements (points) for different spectrometer potentials. Each point contains a time stamp and amplitude for each event in the set.
	\item Events were selected in a specific Si (Li)-detector amplitude window for further processing. We used a fixed wide window that only cuts off electronics noise and didn't affect the signal. All events in the window were counted to get the resulting counting rate at each spectrometer set. This counting rate was also corrected for effective data acquisition dead time of $~8~\mu s$ which also accounted the event pile-up effects.
	\item Then, points at the same spectrometer potential values are averaged over different files for each electron energy set.
\end{enumerate}
		
The integral spectrum after this procedure is shown in Fig.~\ref{fig:scat}
	
\subsection{Extraction of energy loss spectrum shape}
		
The experimental spectrum for energy loss measurements can be represented by the following expression:

\begin{equation}
	\label{equation:convolution}
	S(U) = \int_{U}^{E_{gun}}{R(U,E) \cdot Tr(\varepsilon) dE},
\end{equation}
where $R(U,E)$ is a spectrometer resolution function, see Fig.~\ref{fig:resol} (it is also described in~\cite{aseev2011upper}), $\varepsilon = E_{gun} - E $ is an electron energy loss and $Tr(\varepsilon)$ is a gas target transmission function. This expression also accounts for the electron gun line width by using the resolution function, which was measured with the same electron gun. 
		
\begin{figure}[htb]
	\center
 	\includegraphics[width=0.8\linewidth]{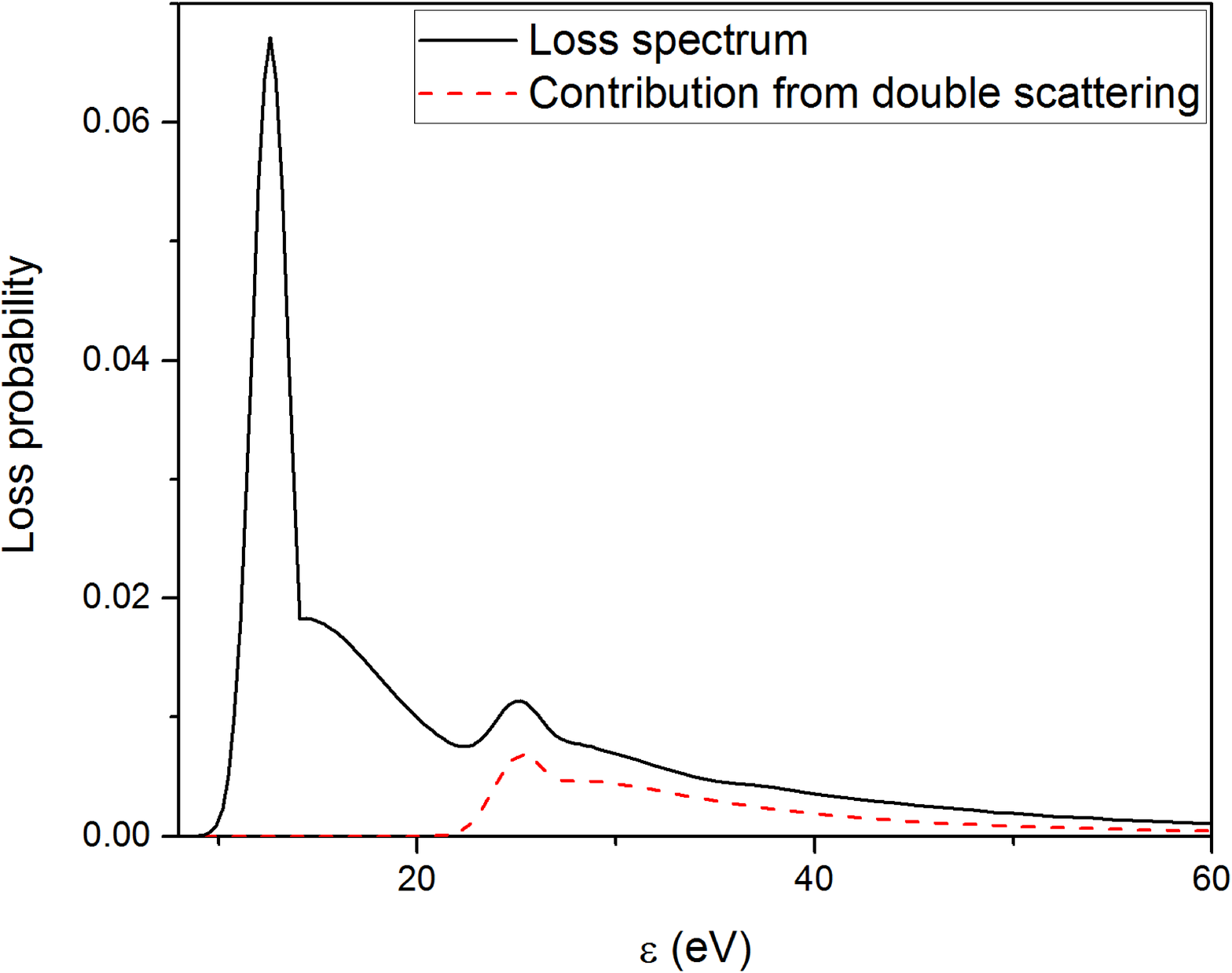}
    \caption{Contribution of double scattering in the measured energy loss spectrum. The triple scattering will be barely seen in this plot.}
	\label{fig:multiple}
\end{figure}

The transmission function $Tr(\varepsilon)$ itself consists of no-loss $\delta(\varepsilon)$ transmission and a sum of single, double, triple and other scattering spectra with appropriate weights $p_i = \frac{X^i e^{-X}}{k!}$. Here $X$ is a dimensionless parameter, called target thickness, which equals the electron inelastic interaction probability during its flight through the target: 
\begin{equation}
X = \frac{l_{target}}{\lambda} = l_{target} n \sigma_{tot}.
	\label{eq:X}
\end{equation}
It allows one to define :
\begin{equation}
	Tr(\varepsilon) = p_0 \delta(\varepsilon) + \sum_i{p_i L_i(\varepsilon)},
\end{equation}
with maximum $i$ automatically chosen so that the probability of the highest order scattering is less than $10^{-4}$.		
The scattering functions $L_i(\varepsilon)$ are obtained by the subsequent convolution of a single scattering loss spectrum $L(\varepsilon) = L_1(\varepsilon)$ with itself, Fig.~\ref{fig:multiple} 

The aim of this experiment is to determine the shape of $L(\varepsilon)$. It is rather difficult to extract the shape of transmission function $Tr(\varepsilon)$ directly from equation \ref{equation:convolution}, but one can define this function and then fit it to the experimental data. 
We used parameterization similar to the one in the previous experiment~\cite{aseev2000energy}:
		
\begin{equation} \label{equation:losses}
	L(\varepsilon) = Norm
	\begin{cases}
		A~ exp \left(  -\frac{2(\varepsilon - P_1)^2}{W_1^2} \right) & \varepsilon \leq \varepsilon_c \\
		\frac{W_2^2}{W_2^2 + 4 (\varepsilon-P_2)^2} & \varepsilon > \varepsilon_c
	\end{cases}
\end{equation}
where $A$ is a constant, $P_i$ and $W_i$ define position and width of the relevant terms, $\varepsilon_c$ is selected dynamically to ensure that the function is continuous and $Norm$ coefficient is calculated automatically to keep the loss probability integral equal to $1$. The first part of this piecewise equation corresponds to the energy losses due to  molecular excitations, the second one describes the continuous ionization spectrum. The distinction of this parametrization from the similar one in~\cite{aseev2000energy} is an automatic normalization and the fact that the position of the excitation peak, $P_1$, was set here as a free parameter. In~\cite{aseev2000energy} it was fixed at 12.6 eV.
		
It should be noted, that since this description is not based on any physical model, the difference between the excitation and ionization parts is rather vague. Moreover, coefficient $A$ does not have any physical meaning due to the dynamic nature of $\varepsilon_c$ and the difference in normalization factors of both parts of the piecewise equation.
		
\subsection{Data fitting and additional calculations \label{section:fitting}}

Data fitting procedure was performed by combination of different fitting algorithms including JMINUIT and Quasi-optimal weights fitter~\cite{aseev2011upper}. Such a complicated procedure is required because of a strong correlation between energy loss function parameters. The spectrometer resolution function $R(U,E)$  in Eq.~\ref{equation:convolution} was taken from the electron gun measurements with an empty WGTS. In addition to the parameters used in Eq.~\ref{equation:losses} the fitting procedure fits the background level, total intensity and the value of $X$.  
		
In order to estimate the ionization part in the total loss spectrum, the following procedure was used: we took analytic loss function with best-fit parameters and calculated the following integral:
		
\begin{equation}
	\label{equation:ratio}
	IonizationTail = \int_{\varepsilon_{low}}^{\infty} L(\varepsilon) d \varepsilon
\end{equation}
		
Such a value does not show the true ratio between excitation and ionization cross-sections, but  $\varepsilon_{low}$ is selected to be larger than any excitation level. That's why this value is proportional to the relative ionization probability. We used $\varepsilon_{low}$ = 17~eV since it guarantees to be above excitation levels.
		
The direct estimation of errors for Eq.~\ref{equation:ratio} is rather complicated, therefore the following procedure was used:
\begin{enumerate}
	\item For each fitted experimental data set there is a vector of the loss function parameters ($A$, $P_1$, $P_2$, $W_1$ and $W_2$) and a corresponding covariance matrix which can be used to construct multivariate normal distribution.
	\item For each such distribution a sample of a few thousand parameter sets is generated.
	\item For each parameter set from this sample a loss function is constructed and integral Eq.~\ref{equation:ratio} is calculated.
	\item The distribution of such integral values of Eq.~\ref{equation:ratio} is considered to be the true distribution of this ratio. It is is proved to be non-normal, but still one can calculate its dispersion and standard deviation for qualitative analysis.
\end{enumerate}

\section{Results}

\subsection{Energy loss function}

\begin{figure}
	\center
	\includegraphics[width=0.7\linewidth]{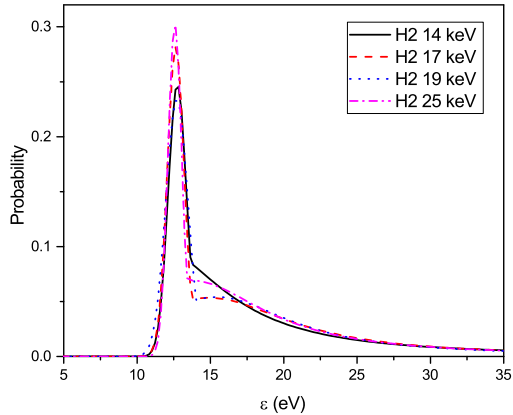}
	\caption{Energy loss spectra $L(\varepsilon)$ at different electron beam energies ($H_2$).}
	\label{fig:H_spectrum}
\end{figure}

\begin{figure}		
	\center
	\includegraphics[width=0.7\linewidth]{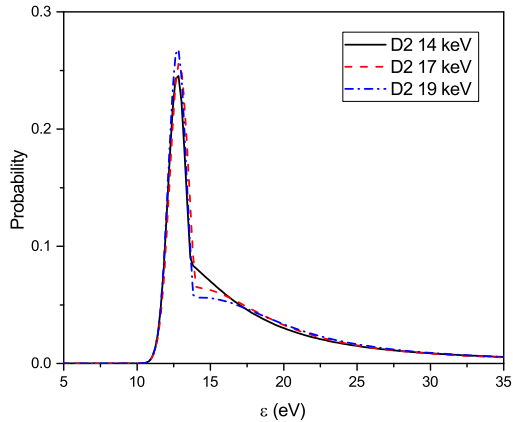}
	\caption{Energy loss spectra $L(\varepsilon)$ at different electron beam energies ($D_2$).}
	\label{fig:D_spectrum}
\end{figure}

\begin{figure}[htb]
\center
 	\includegraphics[width=0.8\linewidth]{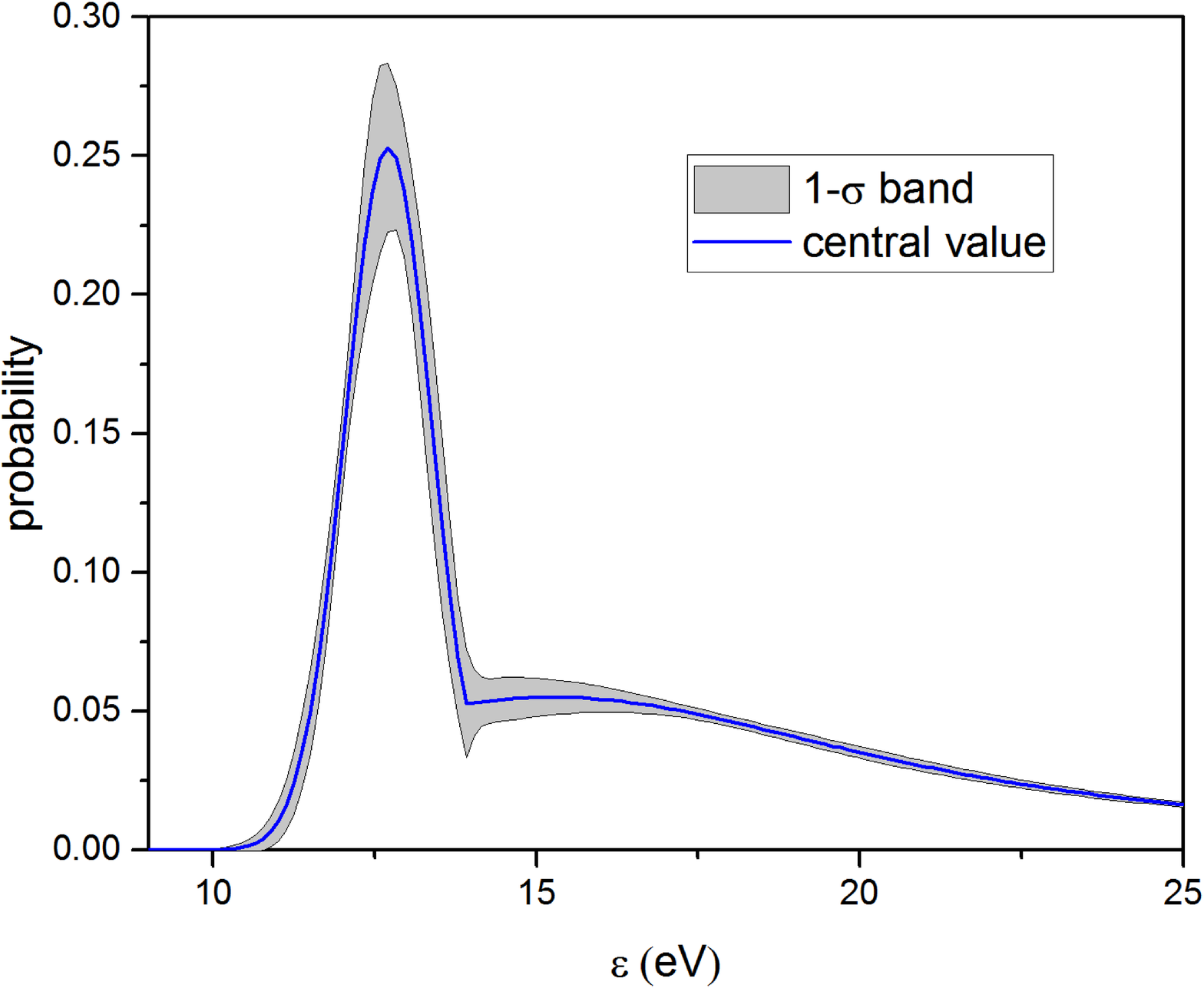}
    \caption{Typical error band for energy loss function obtained by simultaneous randomization of all fitting parameters.}
	\label{fig:loss_error_band}
\end{figure}

The best fits for the energy loss spectra for hydrogen and deuterium at different electron energies are shown in Fig.~\ref{fig:H_spectrum} and Fig.~\ref{fig:D_spectrum}, respectively. Experimental data were fitted by Eq.~\ref{equation:losses} with 5 parameters which are position and width for excitation and ionization peaks and ratio between excitation and ionization (the background, intensity and source thickness were also fitted but were not relevant for the shape analysis). Fig.~\ref{fig:loss_error_band} shows typical error bands for this fitting, which are calculated by the procedure described in Section~\ref{section:fitting}. One may conclude that within our error bars all energy loss functions are the same. For the future practical usage we averaged over all of them. The final fit parameters are presented in Table~\ref{table}. 

The values presented in the table have rather small error bands and it may seem that distinction between parameters for, say, hydrogen and deuterium is rather large in terms of these errors, but one should remember that all the parameters in this parametrization are strongly correlated and, therefore, should not be compared independently. The only visual way to compare such strongly correlated parameter sets is to compare figures like Fig.~\ref{fig:H_spectrum} and Fig.~\ref{fig:D_spectrum} with the error band presented in Fig.~\ref{fig:loss_error_band}.

For the record, this is worth mentioning that ionization potentials for hydrogen atoms and molecules are different. With an accuracy better than $10^{-4}$ the ionization potential for atoms of all hydrogen isotopes is 13.60~eV~\cite{moore}. For molecules these values are 15.426~eV, 15.467~eV, 15.487~eV for H$_2$, D$_2$ and T$_2$, respectively, and can be found in many reference books. Such a precision is beyond the sensitivity of our experiment but it is possible to qualitatively check the similarity between H$_2$ and D$_2$ in our energy range.  

\begin{table}[htb]
	\centering
	\caption{The weighted average fit parameters from Eq.~\ref{equation:losses}.  {\it IonizationTail} describes a portion of ionization in the normalized energy loss function above 17 eV, Eq.~\ref{equation:ratio}. }
	\label{table}
	\begin{tabular}{llllll}
		 Isotope	& $P_1$					& $P_2$					& $W_1$     		& $W_2$				\\
		 D2			& $12.80 \pm 0.04$		& $13.7 \pm 0.5$ 		& $1.31 \pm 0.11$   & $11.62 \pm 0.25$ 	\\
		 H2			& $12.67 \pm 0.02$		& $13.2 \pm 0.2$  		& $1.21 \pm 0.05$	& $12.13 \pm 0.16$	\\
		 average	& $12.695 \pm 0.017$	& $13.29 \pm 0.18$		& $1.22 \pm 0.05$   & $11.99 \pm 0.13$	\\
					& 						& 						& 					& 					\\
		 Isotope	& A						& Norm				& IonizationTail 		\\
		 D2			& $3.66 \pm 0.33$		& $0.068$			& $0.416 \pm 0.004$		\\
		 H2			& $3.59 \pm 0.17$		& $0.070$			& $0.421 \pm 0.002$		\\
		 average	& $3.60 \pm 0.15$		& $0.070$			& $0.420 \pm 0.002$		
	\end{tabular}
\end{table}

\subsection{Relative electron energy dependence}
\begin{figure}[htb]
	\includegraphics[width=0.8\linewidth]{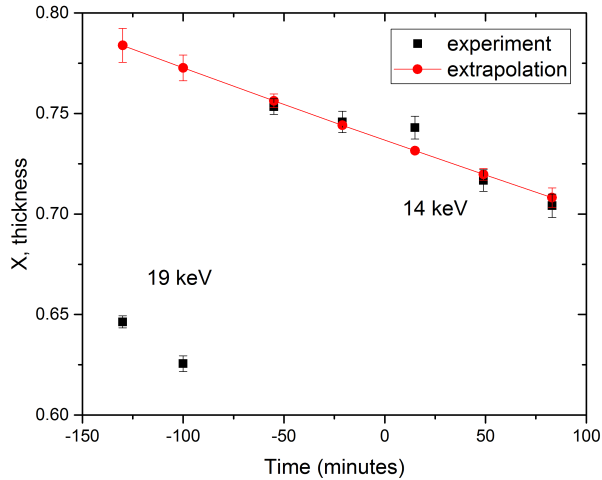}
	\caption{Extracted target thickness during a single deuterium gas fill versus time for two measurements at electron energies of 14~keV and 19~keV -- solid square symbols. Exponential fit to the 14 keV data and its extrapolation to the time of making 19 keV measurements are shown by points and solid line. Error bars show experimental and extrapolation errors.}
	\label{fig:extrapolation}
\end{figure}

The usage of the WGTS as a gas target, in contrast to the solid substrate, makes it rather difficult to compare cross sections at different energies. Because of the slow but continuous hydrogen absorption on the walls of all WGTS components, gas concentration in the pipe goes down with time. Thus, we have to refill WGTS every 5-6 hours and each gas refill has its own particular gas concentration or column density. The typical measurement at one electron energy takes at least an hour, for better statistics -- a few hours. Therefore, some methods should be found to compare electron scattering at different energies with a good precision. We use two methods for such measurements. The first one is to take data at two electron energy during the same gas fill and to make some correction for the change in column density with time. Fig.~\ref{fig:extrapolation} illustrates a slow drop of the extracted gas target thickness $X$ during one deuterium fill. The points for a few runs at 14~keV measurements really show the decrease of $X$ extracted during the fit procedure to experimental data. At the beginning of this particular fill we took two runs at 19 keV as the reference data. The 14~keV points were fitted by exponential function and were extrapolated to the time of 19 keV measurements -- two round symbols with extrapolation error bars on the fitted line. This line represents the relative column density during this fill. One can see that the experimental points at 19~keV are below the extrapolation line giving exactly the difference between the total cross sections at two energies. This relation explicitly comes from Eq.~\ref{eq:X}: $X = l_{target} n \sigma_{tot}$. Such an extrapolation procedure was done for other energy combination 14~keV and 19~keV. We did not take the same fill measurements for 18.7 and 25~keV electrons. 

\begin{figure}[htb]
	\includegraphics[width=0.95\linewidth]{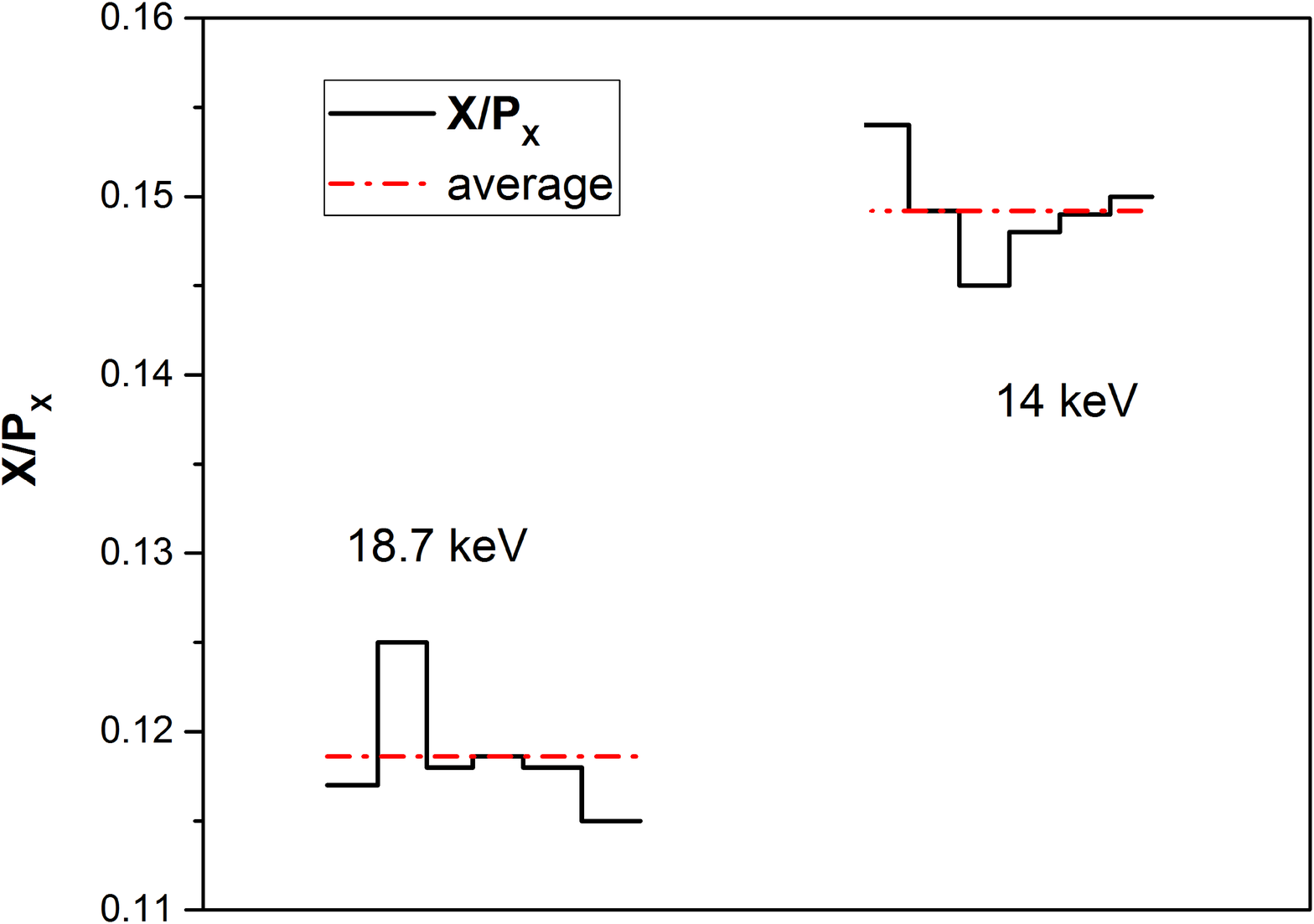}
	\caption{$\frac{X}{P_x}$ ratio during measurements with 18.7~keV and 14~keV electrons. Solid lines correspond to $X$ extracted from the data divided by the pressure sensor $P_x$ value. Dashed dotted line represents the average level.}
	\label{fig:XPx}
\end{figure}

The second  method is aimed to use an empirical fact that gas concentration in the pipe is proportional to the value of our pressure sensor $P_x$ located in the room temperature part of the gas loop, see Section 2. There is a limit on such a measurement determined by precision of $P_x$. The sensor readout has only two digits,  thus at typical values of 2.8--3.2 mm of Hg  the precision is about $0.1/3\approx0.03$. In Fig.~\ref{fig:XPx} we demonstrate stability of $X/P_x$ ratio for two energies. The steps were caused by a change of $X$ and $P_x$ value by $0.1$ from run to run . Difference of $X/P_x$ for two energies reflects the energy dependence of the cross section. At the same column density the cross section at a smaller energy is larger giving a larger extracted value of $X$ at the same proportion.

\begin{figure}[htb]
	\includegraphics[width=0.95\linewidth]{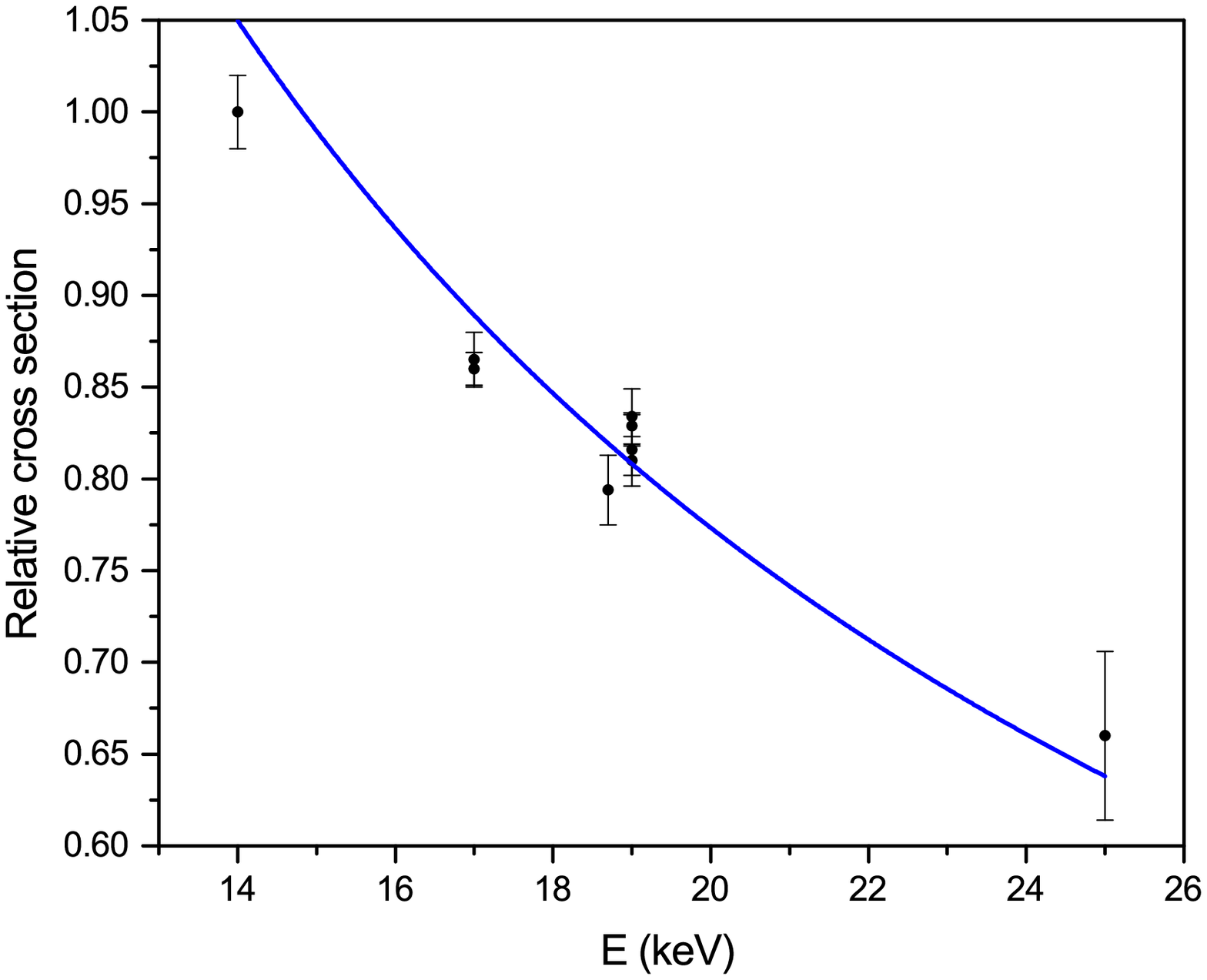}
	\caption{Relative electron scattering cross-section versus the electron energy.  }
	\label{fig:sigma_energy}
\end{figure}

The two methods, extrapolation and $X/P_x$ agree withing the errors where this comparison was possible. 
The resulting relative energy dependence is shown in Fig.~\ref{fig:sigma_energy}. All points were normalized to the value at 14~keV. The solid line represents the fit by function according to Eq.~\ref{eq:energy}. We also averaged values for hydrogen and deuterium which agree within the errors. 

\section{Summary}
The electron scattering on molecular hydrogen and deuterium has been measured at electron energies 14, 17, 18.7, 19 and 25 keV.  Electrons were generated by the electron gun with an energy line width better than 0.3 eV. A windowless gaseous  tritium source (WGTS) of the "Troitsk nu-mass" setup was used as a target. The integral spectrum of scattered electrons was measured by a MAC-E type spectrometer with a relative energy resolution of 8.3$\cdot 10^{-5}$. A phenomenological expression containing excitation and ionization parts was used to fit the integral spectra. As a result, the shapes of molecular excitation and ionization spectra were extracted for both isotopes. There is no difference in extraction spectra between hydrogen and deuterium molecules. The relative energy dependence was extracted too. Two methods were used for this purpose: to take data at two electron energies during one and the same gas fill and to use a pressure sensor $P_x$ for relative normalization of gas density in the pipe. Both methods agree with each other and with the published empirical energy dependence. All the obtained results will be used in data analysis to search for a sterile neutrino in tritium beta decay.

The important result of this experiment is that energy loss spectrum shape and the excitation to ionization ratio do not depend on the incident electron energy in the investigated energy range. This fact will be further used in simulations of the "Troitsk nu-mass" setup.

This work was supported by the Russian Foundation for Basic Research (RFBR) under grant numbers 11-02-00935-a, 12-02-12140-ofim, 14-02-0057015-a and 14-22-0306915-ofim.

\end{document}